\documentclass[final,1p,letterpaper]{elsarticle}

\makeatletter
\def\ps@pprintTitle{%
 \let\@oddhead\@empty
 \let\@evenhead\@empty
 \def\@oddfoot{\centerline{\thepage}}%
 \let\@evenfoot\@oddfoot}
\makeatother

\usepackage{natbib}
\usepackage{cancel}
\usepackage{pifont} 
\usepackage{amsthm}         
\usepackage{amsmath} 	   
\usepackage{amssymb}       
\usepackage{amsfonts}
\usepackage{mathtools}
\usepackage{graphicx} 	    
\usepackage{verbatim}
\usepackage{epsfig,bbm}
\usepackage{color}
\usepackage{colortbl}
\usepackage{float}
\usepackage{multirow}
\usepackage{lipsum}
\usepackage{tensor}
\usepackage{enumitem}

\definecolor{babyblue}{rgb}{0.54, 0.81, 0.94}
\definecolor{corn}{rgb}{0.98, 0.93, 0.36}

\begin{document}
\begin{frontmatter}

\title{Numerical Relativity as a New Tool \\for Fundamental Cosmology}
\author[add1,add2]{Anna Ijjas}
\address[add1]{Center for Cosmology and Particle Physics, Department of Physics,\\ New York University, New York, NY 10003, USA}
\address[add2]{Institute for Theory and Computation, Harvard University, Cambridge, MA 02138, USA}
\ead{ijjas@nyu.edu}

\date{\today}

\begin{abstract}
Advances in our understanding of the origin, evolution and structure of the universe  have long been driven by cosmological perturbation theory, model building and effective field theory. In this review, we introduce numerical relativity as a powerful new complementary tool for Fundamental Cosmology.  To illustrate its power, we discuss applications of numerical relativity to studying the robustness of slow contraction and inflation in homogenizing, isotropizing and flattening the universe beginning from generic unsmooth initial conditions. In particular, we describe how recent numerical relativity studies of slow contraction have revealed a novel, non-linear smoothing mechanism based on {\it ultralocality} that challenges the conventional view on what is required to explain the large-scale homogeneity and isotropy of the observable universe. 
 \end{abstract}

\end{frontmatter}

\section{Introduction} 

Fundamental Cosmology -- the study of the most basic questions underlying the origin, structure and evolution of our universe \cite{Turner:2018bcg} --  owes much of its current success to applying the techniques and tools of high-energy/particle physics to cosmological theory development. This approach rests on the fact that, from the formation of the first elements until the onset of dark energy domination, the relativistic description of our large-scale universe is remarkably simple in two distinct ways:  First, even though our spacetime geometry  as observed by cosmic microwave background (CMB) experiments \cite{Komatsu:2008hk,Planck:2018vyg} is in the non-perturbative regime of Newtonian gravity and hence genuinely relativistic, the essential non-linearity is encapsulated in a single simple dynamical variable, the scale factor $a(t)$ of the flat Friedmann-Robertson-Walker (FRW) line element 
\begin{equation}
\label{FRW}
{\rm d}s^2= - {\rm d}t^2 + a^2(t){\rm d}x_i{\rm d}x^i,
\end{equation}
where the spacetime curvature is given by the Hubble parameter, $H(t) = \dot{a}(t)/a(t)$ with $H(t)$ measuring the  extrinsic 3-curvature $K$ and $H^2(t)$ the intrinsic 4-curvature $R$ of the flat FRW geometry. As per convention,  dot denotes differentiation with respect to the FRW time coordinate, Greek indices refer to spacetime coordinates and Latin indices to refer to purely spatial coordinates. 
Second, the FRW geometry as given by Eq.~\eqref{FRW} is an exact solution of Einstein's field equations,
\begin{equation}
G_{\mu\nu}=T_{\mu\nu},
\end{equation} 
where $G_{\mu\nu}$ is the Einstein tensor and $T_{\mu\nu}$ the stress-energy. Throughout, we use reduced Planck units with $1/M_{\rm PL}^2\equiv 8\pi G_{\rm N}\equiv1$.

The situation is unique. The Einstein equations involve a system of ten coupled, non-linear partial differential equations (PDEs) which are supplemented by the evolution equations for the stress-energy. In general, solving this system of equations is intractable analytically.
Notably, though, evaluated for the flat FRW geometry, the field equations of General Relativity reduce to a set of two simple ordinary differential equations (ODEs), the Friedmann equations,  
\begin{equation}
\label{FRW-eqs}
3H^2 = \rho,\quad -\frac{\dot{H}}{H^2}=\epsilon,
\end{equation}
which describe the evolution of the scale factor $a(t)$ and the stress-energy.  The latter acts on the homogeneous and isotropic FRW background geometry as a barotropic perfect fluid with energy density $\rho$ and pressure $p$ the two of which are related by the equation of state 
\begin{equation}
\label{epsilon-def}
\epsilon\equiv \frac32 \left(1+ \frac{p}{\rho}\right).
\end{equation}

It is important to realize that cosmological physics belongs to the strong-field regime of General Relativity. In particular, the key issues in Fundamental Cosmology are neither in the perturbative regime of flat Minkowski space nor can they be approximated by Newtonian gravity.   
The cases in which the particle theory-inspired perturbative approach to cosmology works in the strong-field regime are the exception.  Several open questions require tools that are capable of handling non-perturbative, non-linear evolution within or beyond Einstein Gravity.
For example, the problem of cosmic initial conditions involves regimes that lie far outside the perturbative FRW geometry. Perhaps the biggest open question in Fundamental Cosmology, the cosmic singularity problem, calls for modifications of Einstein gravity and/or forms of stress-energy that in general lead to non-perturbative effects. 

In this review, we introduce numerical relativity and how it can serve as a powerful complementary tool to further enhance and catalyze advances in Fundamental Cosmology.
The paper is organized as follows, we begin with an overview of the basics of numerical relativity, focusing on its formal and computational aspects that distinguish it from other types of cosmological and astrophysical simulations. We then illustrate the discovery potential of this new tool by discussing recent results that emerged from studying the robustness of early-universe scenarios  to cosmic initial conditions. In particular, we show how non-perturbative studies of slow contraction in Refs.~\cite{Cook:2020oaj,Ijjas:2020dws,Ijjas:2021gkf,Ijjas:2021wml,Ijjas:2021zyf} have found a new way of homogenizing, isotropizing and flattening the universe which is distinct from previously known mechanisms and refutes the standard textbook picture that non-generic, select initial conditions and causal connectedness is necessary for explaining our large-scale universe. We close by comparing to numerical relativity studies of inflation \cite{East:2015ggf,Clough:2016ymm,Clough:2017efm,Aurrekoetxea:2019fhr,Joana:2020rxm} and commenting on future research directions.

\section{Introduction to Numerical Relativity}

For many decades, sophisticated computer simulations have been a common and heavily utilized tool in cosmology. Whether it comes to understanding the cosmic microwave background or structure formation, computation has been  indispensable in evaluating  existing theoretical models in light of observational data. In this Section, we introduce numerical relativity as a novel kind of computational tool for cosmology with the simulation not being an end point of the investigation but, rather, an intermediate discovery and exploration tool for theory development.

There are numerous introductions to numerical relativity mostly focused on simulating compact object mergers (see, {\it e.g.}, \cite{Baumgarte:2021skc}). Here, we do not intend to replace or repeat existing overviews. Instead, in presenting the very basics of numerical relativity, we will focus on two aspects as they relate to Fundamental Cosmology:
\begin{itemize}
\item the necessary `formal dressing' of the field equations so they are suitable for numerical integration; and 
\item the basic structure of the computation involving specifying initial data, numerical integration of the PDE system as well as code validation. 
\end{itemize}

\subsection{Formal dressing of the field equations}
\label{subsec:WP}

The diffeomorphism invariance or gauge freedom of General Relativity poses a non-negligible problem for computation in the sense that most formulations of the field equations are ill-suited for numerical integration.  Here, the term `formulation' for a given theory as established in mathematical relativity ({\it e.g.}, see \cite{Geroch-IVP}) is defined to be the representation of the underlying system of differential equations obtained by choosing a particular coordinate system. Note that a formulation is not equivalent to a gauge choice. For example, different gauge conditions can be implemented in the same formulation.
A `good' formulation for numerical applications is one that leads to a `well-posed' initial value problem which is `well-suited' to the underlying physical situation.

A formulation is called `well-posed'  if  the underlying system of differential equations can be put into a form such that, for given initial conditions, there exists a unique solution that continuously depends on the initial conditions.  The challenge of finding a well-posed formulation of the field equations is beyond what we typically encounter in Fundamental Cosmology. In perturbative analyses, the gauge choice is a matter of convenience and simplicity while the issue of formulation hardly ever arises. That is, in particle physics-inspired approaches to cosmology, the underlying formulation of the Einstein equations is almost always the coordinate-based Arnowitt-Deser-Misner (ADM) slicing which is a foliation with spacelike 3-hypersurfaces  evolving in time. At each perturbative order, this slicing is supplemented by {\it algebraic} gauge conditions such that the lapse and the three components of the shift vector define a preferred timelike congruence.  

The ADM formulation of the field equations is a natural and appropriate choice in the particle theory/EFT context. In particular, utilizing the symmetries of the flat FRW background solution, the ADM form greatly simplifies the perturbative analysis, as it enables the separation of linear perturbations around this background into decoupled scalar, vector, and tensor degrees of freedom that each evolve independently mode by mode \cite{Bardeen:1980kt}.  Instead of an analytically intractable non-linear system of ten plus coupled PDEs, it suffices to consider a handful of decoupled linear ODEs. The power of cosmological perturbation theory to track the dynamics as well as to extract predictions for CMB and other observations has been demonstrated numerous times over the course of the past four decades (for a comprehensive summary from the gravitational perspective, see {\it e.g.,} \cite{Ijjas:2018cdm}). 

However, the ADM form is ill-suited when it comes to questions beyond perturbation theory, especially those which can only be addressed by numerical integration of the field equations. The reason is that an ADM formulation of the field equations of General Relativity supplemented by algebraic gauge conditions yields a weakly-hyperbolic PDE system and, hence, an ill-posed initial value problem.

An important example of a well-posed formulation is the coordinate-based generalized `harmonic' form of the field equations. The defining feature of this formulation is that each of the spacetime coordinates $x^{\mu}$ obeys a scalar wave equation with source function $J^{\mu}$ that itself is a function of the coordinates,
\begin{equation}
\label{harm-coo-def2}
\Box x^{\mu} = J^{\mu}(x^{\alpha})\,.
\end{equation}
Here, the source functions $J^\mu$ are treated as additional degrees of freedom, with Eq.~\eqref{harm-coo-def2} yielding a set of constraint equations. By choosing appropriate source functions, any known metric $g_{\mu\nu}$ in {\it any} coordinate system can be expressed in generalized harmonic form.  In harmonic form, the principal part of Einstein's field equations is particularly elegant, yielding a wave equation for each metric tensor component.  Together the equations constitute a strongly-hyperbolic PDE system which is manifestly well-posed.
Notably, the harmonic formulation was utilized for both the first local, non-linear well-posedness proof of the field equations of General Relativity (in vacuum) \cite{FouresBruhat:1952zz} as well as the first successful numerical integration of black hole mergers \cite{Pretorius:2004jg}. 

A well-posed formulation is `well-suited' to the physical situation if it addresses issues specifically related to the dynamics and/or geometry under consideration. For example, a well-posed formulation which is well-suited to simulate compact object mergers is ill-suited to simulate cosmological spacetimes. In particular, for slowly contracting spacetimes as considered in this review,  we must address a stiffness problem to accurately track contraction over several hundreds of $e$-folds. 
The stiffness problem arises because the equation of state parameter $\epsilon$ as defined in Eq.~\eqref{epsilon-def} is much greater than three which means that  the the scale factor $a(t)$ is decreasing very slowly compared to the Hubble radius $|H^{-1}| \propto a^{\epsilon}$.  
This stiffness problem has been successfully addressed by using a `Hubble-normalized' formulation in which the Hubble radius does not enter explicitly (see {\it e.g.} \cite{Garfinkle:2008ei,Ijjas:2021gkf}). The analogous problem pertaining to studying the robustness of inflationary spacetimes to cosmic initial conditions is yet unresolved, as we will discuss below in Sec.~\ref{subsec:inflation}. 

While well-suitedness is an intuitively obvious requirement, well-posedness might appear an abstract and unnecessary complication. However, a well-behaved numerical relativity computation  demands a well-posed formulation. For example, given some initial data, an ill-posed formulation might lead to no or multiple solutions. A common manifestation of an ill-posed formulation is a `blow-up' within finite time even in the absence of any physical instability, where the blow-up time is resolution dependent. As a result, one cannot establish convergence and, consequently, cannot  reach a valid numerical solution.

\subsection{Basic structure of Numerical Relativity codes}
\label{sec:NR}

Cosmological numerical relativity codes that involve time-dependent PDE systems have three non-trivial components: 
\begin{itemize}
\item specifying initial conditions, 
\item evolving the PDE system, and
\item code validation.
\end{itemize}

In General Relativity, setting the initial data, which is then evolved by numerically solving the Einstein PDE system, means specifying the {\it geometric} variables that define the 3-metric $\gamma_{ij}$ and the extrinsic 3-curvature $K_{ij}$ of a spatial hypersurface $\Sigma_0$ at some initial time $t_0$ as well as specifying the {\it stress-energy} variables that describe the energy and momentum densities, $\rho$ and $p_j$, as well as the anisotropic stress $\Pi_{ij}$ as measured by the Eulerian observer at time $t_0$. 

The initial data cannot be chosen arbitrarily.  The reason is that the Hamiltonian and momentum constraints, 
{\it i.e.}, the Einstein equations projected orthogonally onto the $t_0$-hypersurface $\Sigma_0$,
\begin{eqnarray}
\label{Cg}
{}^{(3)}R + K^2 - K_{ij}K^{ij} &=& 2\rho,\\
\label{Cc}
D_iK^i{}_j - D_jK&=& p_j,
\end{eqnarray}
where $D_i$ denotes the covariant derivative associated with the 3-metric $\gamma_{ij}$,
 are {\it not} satisfied for generic initial data. 
The field equations can be evolved using unphysical initial data that violate energy and momentum conservation. But, on the other hand, if the initial data are chosen to be constraint satisfying, the field equations of General Relativity propagate and preserve the constraints going forward in time.
Hence, the first non-trivial component of a numerical relativity code is to ensure that the geometric and stress-energy variables, $\gamma_{ij}, K_{ij}, \rho$ and $p_j$, satisfy the Hamiltonian and momentum constraints~(\ref{Cg}-\ref{Cc}). In practice, this means freely choosing some variables and then numerically solving the Hamiltonian and momentum constraints~(\ref{Cg}-\ref{Cc}), which corresponds to solving four coupled elliptic equations to specify the remaining components. This is an essential computational step to explore generic initial conditions and reach general conclusions.

Once constraint satisfying initial data is specified, the data is evolved by numerically solving the Einstein field equations. The underlying numerical scheme must be a well-posed formulation which is well-suited to the physical situation, as emphasized above.  The computational tools and techniques (for an introductory overview see, {\it e.g.}, \cite{Lehner:2001wq}) are common to other situations involving PDE systems. 
However,  in addition to the usual sort of convergence tests required in numerically solving PDEs, numerical relativity demands special validation tests associated with the constraints.

Free evolution schemes, such as those underlying the simulations discussed in this review, numerically integrate  the dynamical PDE system but only use the constraints in specifying the initial data. Since the Hamiltonian and momentum constraints can only be satisfied initially up to some numerical error, estimating how the truncation error evolves with time is essential to verify physically consistent convergence of the code.  For example, to establish convergence, one might plot the Euclidean norm of the Hamiltonian constraint integrated over the spatial domain as a function of time, computed  at three (or more) different resolutions (see, {\it e.g.}, Appendix~A in Ref.~\cite{Ijjas:2020dws}). With increased resolution, the truncation error decreases in a predictable way. The code passes the convergence test if, after appropriate rescaling, the two curves computed at finer grained resolutions lie proportionately closer to one another than the two curves corresponding to coarser grained resolutions.

Finally, a well-posed formulation must be supplemented with an effective constraint damping scheme if the unphysical truncation error happens to grow exponentially. To prevent such an artificial blow-up, terms are added to the evolution equations that damp the growth of constraint violating numerical errors, while leaving the underlying solution unaffected in the continuum limit \cite{Gundlach:2005eh,Brodbeck:1998az,Pretorius:2005gq,Pretorius:2006tp}.

\section{Smoothing through Ultralocality}

To illustrate the discovery potential of numerical relativity as a tool in Fundamental Cosmology, we discuss in this Section recent numerical relativity studies of two early-universe scenarios, slow contraction and accelerated expansion (inflation).  
The aim of both types of studies was to test the sensitivity of the dynamics to cosmic initial conditions. As originally imagined, both slow contraction and inflation are supposed to homogenize, isotropize and flatten (henceforth, {\it smooth}) the universe.  We refer to a smoothing mechanism as {\it robust} if it homogenizes, isotropizes and flattens for a wide range of initial conditions including those outside the perturbative regime of FRW spacetimes. 

Whether a smoothing mechanism is robust is a central issue in cosmology because homogeneity, isotropy and flatness are highly non-generic conditions in General Relativity.  Yet all the successful predictions of astrophysical cosmology (nucleosynthesis, recombination and the cosmic microwave background, structure formation beginning from small density perturbations, Hubble expansion, {\it etc.}) rely on the assumption that some process occurred that transformed an inhomogeneous, curved and anisotropic early universe into the smooth universe we observe today.  

In the case of inflation, robustness studies \cite{East:2015ggf,Clough:2016ymm,Clough:2017efm,Aurrekoetxea:2019fhr,Joana:2020rxm} are (yet) restricted to a set of special initial conditions that are known to favor inflation.
 In the case of slow contraction, the results demonstrate remarkably robust smoothing  \cite{Garfinkle:2008ei,Cook:2020oaj,Ijjas:2020dws,Ijjas:2021gkf,Ijjas:2021wml}. 
 However, the more profound lesson from the slow contraction studies on which we will focus in this Section is the discovery that smoothing can occur in a fundamentally different way than conventionally assumed based on perturbative analyses.

\subsection{Conventional picture of Smoothing}
\label{sec:standard}

The standard textbook explanation for how the universe can be smoothed, whether through inflation or slow contraction, does not address the issue of robustness.  Instead, the standard argument assumes the existence of a rare, select patch of spacetime greater than the initial Hubble volume that is nearly homogeneous and isotropic to begin with and then explains how smoothness can be reached over exponentially many Hubble volumes. A Hubble volume is a patch whose radius is of order $c |H|^{-1}$, where $c$ is the speed of light.

In this approach the evolution of the select patch is well-described by the generalized Friedmann equation:
\begin{equation}
\label{gen-FRW-constraint}
3H^2 = \frac{\rho_m}{a^3} + \frac{\rho_r}{a^4}  - \frac{k^2}{a^2} + \frac{\sigma^2}{a^6} + \frac{\rho_{\phi}}{a^{2\epsilon}} . 
\end{equation}
Here $\rho_m, \rho_r, k$ and $\sigma$ are constants that characterize  the individual contributions of pressureless matter, radiation, (zero-mode) spatial curvature and anisotropy to the total energy density at some initial time $t_0$. The parameter $ \epsilon$ denotes the equation of state of the `smoothing' stress-energy component which is commonly assumed to be sourced by one or more scalar fields and $\rho_{\phi}$ is the sum of the field's kinetic and potential energy density at $t_0$. 

In an expanding universe ($\dot{a}>0$), the total energy density, which is proportional to $H^2$, as well as all individual contributions on the r.h.s. of Eq.~\eqref{gen-FRW-constraint} decrease with growing scale factor $a$. But the components decrease at different rates such that the one that decreases the slowest eventually comes to dominate the total energy density. 
In order for the smoothing component to dominate the  anisotropy ($\propto 1/a^6$) and curvature ($\propto 1/a^2$),  its equation of state $\epsilon$ must be less than one. A phase with $\epsilon<1$ corresponds to accelerated expansion or inflation \cite{Guth:1980zm,Linde:1981mu,Albrecht:1982wi}, $\ddot{a}\propto 1-\epsilon>0$. 
Furthermore, the smoothness of the initial patch is extended over exponentially many Hubble volumes as the patch ($\sim a^3$) rapidly grows compared to the Hubble volume ($\sim H^{-3} \propto a^{6\epsilon}$).

Similar logic was used when slow contraction was proposed as an alternative smoothing mechanism to inflation \cite{Khoury:2001bz,Erickson:2003zm}. Here, the only difference is due to the fact that, in a contracting universe ($\dot{a}<0$), all individual contributions on the r.h.s. of Eq.~\eqref{gen-FRW-constraint} increase with shrinking scale factor $a$. Therefore, to dominate anisotropy and curvature,  the energy density of the smoothing component must grow at the fastest rate. This requires $\epsilon>3$. At the onset of slow contraction, the Hubble radius of the select patch is exponentially large compared to the case of inflation because the energy density within the patch is tiny compared to the energy density at the onset of inflation.
By the end of slow contraction, the smoothness of the initial patch extends over exponentially many Hubble volumes simply because the Hubble radius shrinks while the physical size of the patch does not \cite{Ijjas:2019pyf,Cook:2020oaj}.

According to the standard textbook picture, then, smoothing requires:
\begin{itemize}
\item {\it Non-generic, select initial conditions}. The mechanism of inflation or slow contraction can only start if there exists a patch of sufficient size (Hubble volume or larger) with only small deviations from homogeneity and isotropy and the patch is dominated by a homogeneous scalar field with special potential energy density, initial field value and velocity to achieve the necessary value of $\epsilon$.
\item {\it Causal connectedness}. Homogeneity and isotropy over exponentially many Hubble volumes originate from the assumed homogeneity and isotropy of the initial patch. Generic inhomogeneities outside the initial patch do not get smoothed out; they only become inaccessible to the local observer within the initial patch. In other words, smoothing only occurs in select patches whose conditions obviously favor inflation or slow contraction.
\end{itemize}

In recent years, advances in numerical relativity have made it possible to study whether smoothing occurs under generic initial conditions, {\it i.e.},  outside the perturbative regime of FRW spacetimes. These studies are important because they test whether the above principles properly represent what is required to smooth generally.

\subsection{Smoothing, Robustness and Ultralocality}
\label{sec:ultralocal}

In a series of recent studies \cite{Cook:2020oaj,Ijjas:2020dws,Ijjas:2021gkf,Ijjas:2021wml}, the extraordinary robustness of slow contraction has been established using the tools and techniques of mathematical and numerical relativity. The analyses involved several hundred simulations that each evolved the non-linear Einstein-scalar PDE system under a distinct set of initial conditions and with different shapes for the scalar field potential. As a key finding, the studies demonstrate that a long period of smoothing slow contraction occurs for a wide range of initial conditions including those that lie far outside the perturbative regime of FRW spacetimes. Impressively, smoothing is observed even under conditions that strongly disfavor the onset of slow contraction: large non-linear inhomogeneous anisotropy and spatial curvature contributions with some regions expanding initially as well as wild field velocity distributions including regions in which the scalar field rolls uphill initially. 
The combined result is important because it refutes the standard textbook picture in which smoothing slow contraction only occurs under select favorable initial conditions.

More surprisingly, though, the robustness studies also defy the idea that smoothing slow contraction requires causal connectedness. With no exceptions, the many hundred simulations have revealed that smoothing during slow contraction is a two-stage process: First,  the geometry rapidly converges to an inhomogeneous, spatially-curved and anisotropic `ultralocal' state. A state is called ultralocal if the spatial gradient contribution, measured relative to parallel transported coordinates, is negligibly small compared to the contribution by pure time derivatives. Having reached the ultralocal state, then, each spacetime point independently converges to a smooth  FRW spacetime. 

The concept of ultralocality dates back nearly half a century although in a different context and {\it without} reference to smoothing. Originally, Belinski, Khalatnikov and Lifshitz conjectured in Ref.~\cite{Belinsky:1970ew} that, in contracting vacuum space-times, spatial gradients in the equations of motion become small compared to the time derivatives. Several numerical analyses studying how relativistic space-times approach a putative singularity provide evidence for the conjecture in some special settings, assuming certain symmetry conditions or a particular matter source (vacuum, stiff fluid, or a free scalar) \cite{Berger:1998vxa,Lim:2009dg,Garfinkle:2020lhb}.  

\begin{figure*}[tb]
\begin{center}
\includegraphics[width=\textwidth,angle=-0]{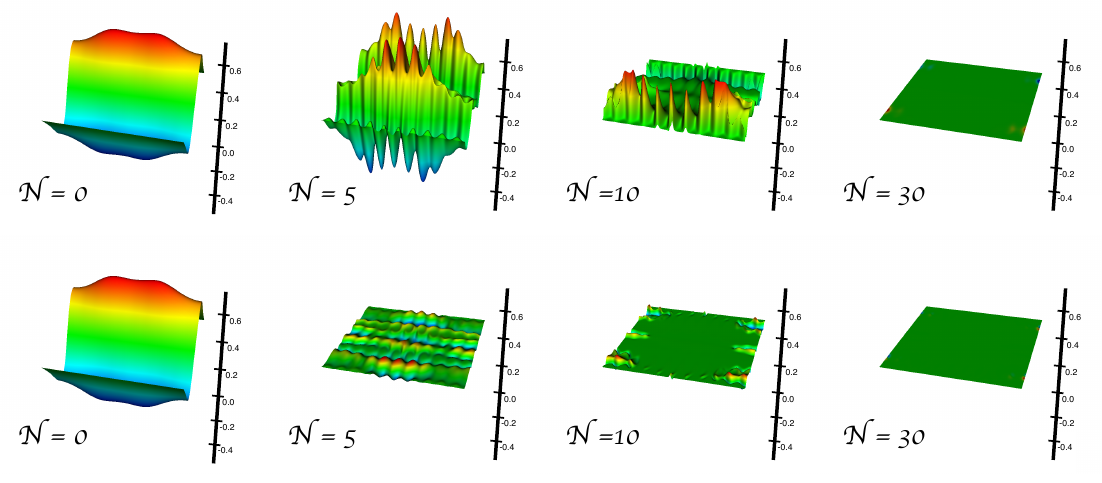}
\end{center}
\caption{Four ${\cal N}=$ const. snapshots of the diagonal $11$-component $\bar{n}_{11}$ (upper panel) of the intrinsic (spatial) curvature tensor $\bar{n}_{ab}$ corresponding to the simulation first presented in Ref.~\cite{Ijjas:2021gkf} with the initial average velocity $Q_0=0.1$. The lower panel shows the pure gradient contribution to $\bar{n}_{11}$. The time coordinate ${\cal N}$ marks the number of $e$-foldings of contraction in the Hubble radius $H^{-1}$. Notably, complete smoothing of the initial patch is reached by only 30 $e$-folds of contraction.
} 
\label{Fig_n11}
\end{figure*} 

To illustrate the role ultralocality plays in smoothing, we follow the evolution of the Hubble-normalized intrinsic (spatial) 3-curvature tensor $\bar{n}_{ab}$ as a function of Hubble time ${\cal N}$ from a representative simulation that starts with non-perturbative initial conditions and converges to the stable flat FRW attractor solution. (For details of the underlying numerical scheme and specific initial conditions as well as a comprehensive description of the numerical results see Ref.~\cite{Ijjas:2021gkf}.) 
The upper panel of Figure~\ref{Fig_n11}  shows the evolution of the diagonal component $\bar{n}_{11}$ at four different time steps. The lower panel shows the  spatial gradient contribution only. Due to Hubble-normalization, all quantities are dimensionless and only take values ranging from $-1$ to $+1$. The overall box size, which corresponds to the initial patch, has edge length $2\pi|H_0^{-1}|$. Its physical size does not change measurably during $30$ $e$-foldings of slow contraction.
The large inhomogeneous curvature component at ${\cal N}=0$ in the upper panel indicates that the initial state is far outside the perturbative regime of flat FRW spacetimes. Comparing the values from upper and lower panel at ${\cal N}=0$, it is apparent that the gradient contribution is significant initially. This changes within only a few $e$-foldings of contraction: At ${\cal N}=5$, the gradient contribution to $\bar{n}_{11}$ decreased  exponentially (the ultralocality effect) while the homogeneous spatial curvature contribution further increased.
By ${\cal N}=10$, the state is nearly ultralocal, {\it i.e.}, the spatial curvature is still very large almost everywhere but most of the gradient contribution decayed to negligibly small values. Complete smoothing is reached within only ${\cal N}=30$ $e$-foldings, corresponding to a decrease in $|H^{-1}|$ by a factor of $e^{30}$.
 \begin{figure}[tb]
\begin{center}
\includegraphics[width=3.25in,angle=-0]{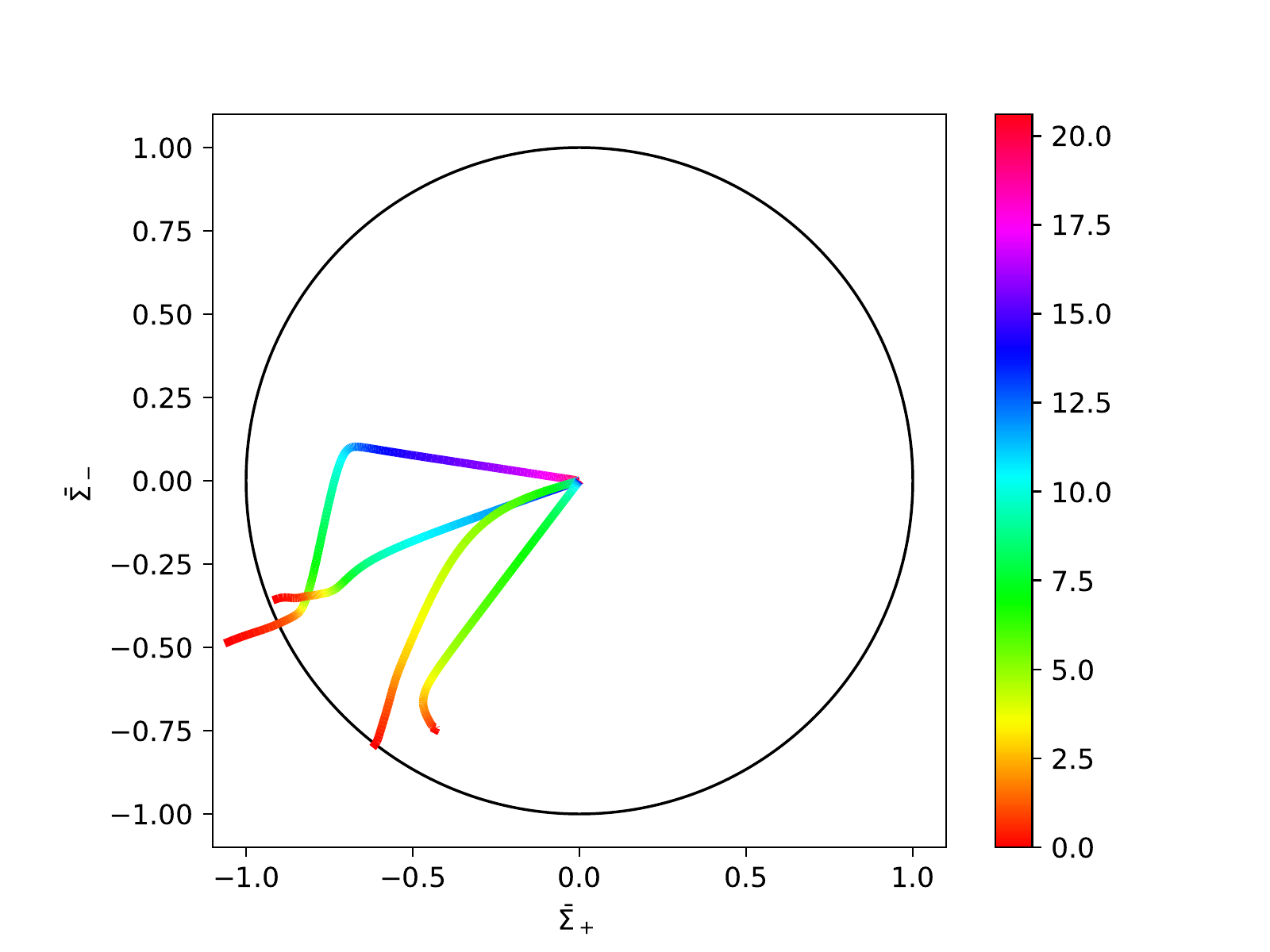}
\end{center}
\caption{The state space orbit for a worldline at four different spatial points. Each point is located in one of the initially inhomogeneous regions shown in Fig.~\ref{Fig_n11}. The circle corresponds to the (vacuum) Kasner solution and the center of the circle corresponds to a smooth FRW universe. The color coding along each of the four state space orbits indicates the number of $e$-folds of contraction of the Hubble radius $|H^{-1}|$ according to the color bar on the right. Although the four points start with different initial conditions and follow different trajectories that reach the center at different times, all converge to the flat FRW state after only 20 $e$-folds. 
\label{CirclePlot}}
\end{figure}  

To illustrate the evolution of the Hubble-normalized extrinsic 3-curvature (shear or anisotropy) tensor $\bar{\Sigma}_{ab}$, in Figure~\ref{CirclePlot} we show state space orbit plots in the $(\bar{\Sigma}_+, \bar{\Sigma}_-)$ plane for four distinct spacetime points from the initial patch of Fig.~\ref{Fig_n11}, where 
\begin{equation}
\label{Sigmaplus}
\bar{\Sigma}_+ = \textstyle{\frac12}\Big(\bar{\Sigma}_{11} + \bar{\Sigma}_{22}\Big)
,\quad
\bar{\Sigma}_- = \textstyle{\frac{1}{2\sqrt{3}}}\Big(\bar{\Sigma}_{11} - \bar{\Sigma}_{22}\Big).
\end{equation}
The $\bar{\Sigma}_{\pm}$ are normalized such that the unit circle ($\bar{\Sigma}_+^2+\bar{\Sigma}_-^2 =1$) corresponds to the vacuum Kasner solution. The center of the circle to the flat FRW solution.
The orbits begin near the outer (Kasner) circle and travel inward. They demonstrate that each of the four points reaches the flat FRW state yet traverses very different trajectories with the fastest smoothing occurring within $\sim 5$ $e$-foldings and the slowest within $\sim 18$ $e$-foldings.

The results, which are representative of hundreds of numerical relativity simulations, dramatically change the standard textbook picture  of smoothing through slow contraction:
\begin{itemize}[leftmargin=*]
\item First,  the overall Hubble-normalized spatial curvature contribution does not decay before the overall Hubble-normalized anisotropy contribution decays. Instead, the sequence is that the gradient contributions to the spatial curvature and shear decay first, before the homogeneous spatial curvature and shear contributions decay.
\item Second, smoothing does not require causal connectedness. In fact, regions of the inhomogeneous, anisotropic and spatially-curved initial patch become causally disconnected well before they reach the smooth FRW state, as the Hubble radius shrinks rapidly while the scale factor and, hence, the initial patch hardly shrinks at all. In the example of Fig.~\ref{Fig_n11}, by the time the initial patch reaches the spatially-curved and anisotropic ultralocal state at ${\cal N}=10$, it consists of  $\sim e^{30} \approx 10^{13}$ causally disconnected and not yet smooth Hubble volumes that each individually converge to the flat FRW state by ${\cal N}=30$. 
\item Third, smoothing is {\it universal}. That is, smoothing is not restricted to regions that are initially nearly homogeneous and isotropic. Instead, as illustrated in Fig~\ref{Fig_n11}, spacetime points of the inhomogeneous, anisotropic and spatially-curved initial patch independently converges to the smooth FRW state.
%
\begin{figure*}[tbp]
\begin{center}
\includegraphics[width=2.5in,angle=-0]{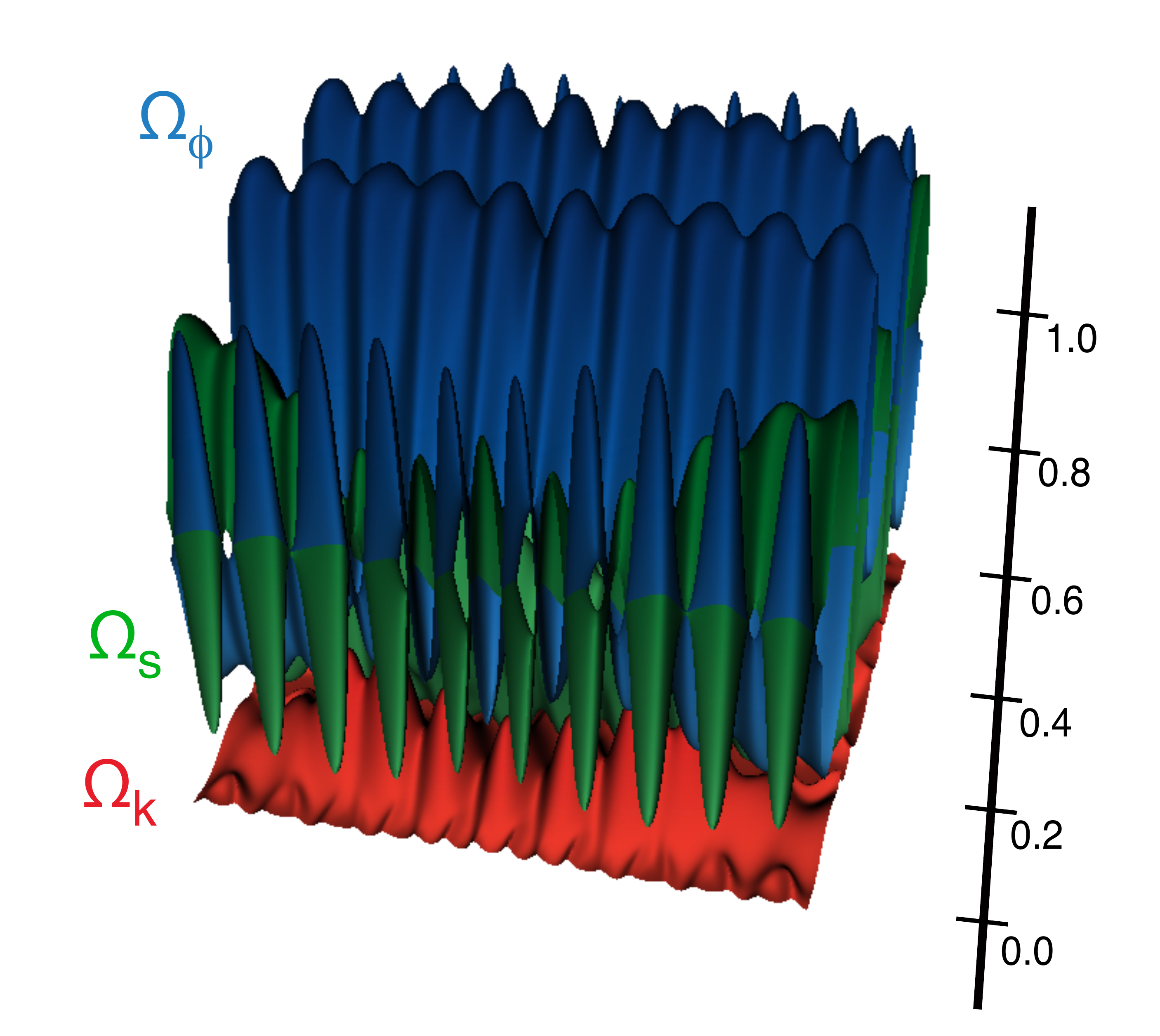}
\end{center}
\caption{Snapshot of the relative contributions of the three energy density components, $\Omega_{\phi}=$~scalar field (blue), $\Omega_k=$~spatial curvature (red) and $\Omega_s=$~shear (green) at ${\cal N}=5$. As Fig.~\ref{Fig_n11} illustrates, the state is ultralocal. Yet, as observed above, in many regions  shear  (green) dominates and the spatial curvature contribution (red) is non-negligble. 
 }    
\label{Fig_Omega3}
\end{figure*} 
\item Fourth, the scalar field need not dominate the energy density well before reaching the flat FRW state. Instead, together with the other gradient contributions, the scalar field spatial gradients decay because of the ultralocality of contraction. As illustrated in Fig.~\ref{Fig_Omega3}, the ultralocal state is generically spatially-curved and anisotropic with the scalar field energy density being sub-dominant in many regions. The flat FRW state is reached because a canonical scalar field with a negative potential energy density destabilizes the anisotropy-dominated Kasner state which is the only other fixed point of the ultralocal state (see, Sec.~5 in Ref.~\cite{Ijjas:2020dws} for an analytic proof). The spatially-curved and anisotropic ultralocal state is then forced to evolve to the smooth FRW attractor solution. 
\end{itemize}

\subsection{Comparison to robustness studies of Inflation}
\label{subsec:inflation}

In the remainder of this section, we will discuss recent numerical relativity studies of inflation \cite{East:2015ggf,Clough:2016ymm,Clough:2017efm,Aurrekoetxea:2019fhr,Joana:2020rxm}. Our focus will be how they  compare to the slow contraction studies \cite{Garfinkle:2008ei,Cook:2020oaj,Ijjas:2020dws,Ijjas:2021gkf,Ijjas:2021wml} in establishing robustness, which, as we have emphasized, is an essential feature for any successful smoothing mechanism. That is, 
\begin{itemize}
\item how does the range of initial conditions compare in numerical relativity studies of slow contraction and inflation?;  and
\item how do the numerical relativity studies of slow contraction and inflation each address the fact that the problem has two fundamental length scales -- corresponding to the physical volume and the Hubble volume -- which evolve at exponentially different rates? 
\end{itemize}
To establish robust smoothing, the underlying mechanism must be insensitive to a wide range of initial conditions;  and it must last sufficiently long to smooth a super-Hubble volume that will later encompass today's observable universe, while maintaining smoothness on the scale of the Hubble volume. Only demonstrating a short period of smoothing for a select, non-generic set of initial conditions is insufficient to establish robustness.
 
{\it Initial conditions.} As detailed above in Sec.~\ref{sec:NR}, in numerical relativity studies, setting the initial conditions means specifying the 3-metric $\gamma_{ij}$ and the extrinsic 3-curvature $K_{ij}$ of a spatial hypersurface at some initial time $t_0$ as well as a full set of variables that fully specify the stress-energy at $t_0$ while obeying the Hamiltonian and momentum constraints~(\ref{Cg}-\ref{Cc}). If the stress-energy is sourced by a canonical scalar field, as in all robustness studies considered in this review, specifying the stress-energy means fixing the field $\phi$ and its velocity $\dot{\phi}$ at $t_0$.
 
A standard tool which is commonly employed in the numerical relativity literature is York's conformal method \cite{York:1971hw}. In most applications, the initial 3-metric is fixed to be conformally-flat, {\it i.e.}, 
\begin{equation}
\gamma_{ij}(t_0, {\bf x}) = \psi^4(t_0, {\bf x})\delta_{ij}.
\end{equation}
Note that generality is not lost by choosing a conformally-flat initial 3-metric. This is simply a device to ensure constraint satisfying initial conditions. However, while the initial constraint satisfaction propagates forward in time, as required, the initial conformal flatness is strongly broken in just a few integration steps, as shown, {\it e.g.}, in the case of slow contraction \cite{Ijjas:2021gkf,Ijjas:2021wml}.
When fully exploited, York's conformal method enables freely specifying the initial conditions through a maximum number of variables while satisfying the Hamiltonian and momentum constraints~(\ref{Cg}-\ref{Cc}).

For example, in numerical relativity studies of slow contraction, the scalar field velocity as well as the shear tensor, which is the trace-free part of the extrinsic 3-curvature $K_{ab}$, are rescaled so the constraints~(\ref{Cg}-\ref{Cc}) decouple,  
\begin{eqnarray}
\label{Cg-conf}
\partial^i \partial _i \psi &=& {\textstyle \frac14 } \left( {\textstyle \frac13}K^2 - V  \right) \psi^5 - {\textstyle \frac18 } \left( \partial^i \phi \partial _i \phi \right)\psi
- {\textstyle \frac18 } \left( Q^2  + Z^{ab} Z_{ab} \right) \psi^{-7},\\
\label{Cc-conf}
D^aZ_{ab} &=& Q D_b\phi(t_0, {\bf x}).
\end{eqnarray}
Evaluated for the rescaled variables, the momentum constraint enables us to freely specify the initial divergence-free (transverse traceless) part $Z_{ab}^{TT}(t_0, {\bf x})$ of the shear tensor  $Z_{ab}(t_0, {\bf x})$, the scalar field distribution $\phi(t_0, {\bf x})$ as well as its velocity $Q(t_0, {\bf x})$ such that the initial conditions space has 30 independent parameters.  The full set of initial conditions considered in slow contraction studies is detailed in Table~\ref{table-1}.
%
\begin{table*}[tb]
\centering
\renewcommand{\arraystretch}{1.75}
\begin{tabular}{ |c| c| }
 \hline
 Ref.  & Garfinkle et al~\cite{Garfinkle:2008ei}; Cook et al~\cite{Cook:2020oaj}; Ijjas et al~\cite{Ijjas:2020dws,Ijjas:2021gkf,Ijjas:2021wml}
  \\ \hline  \hline
$\phi(t_0, {\bf x})$    & $\phi_0 + \sum\limits_{i=1}^{3} f_{i} \cos{(n_i x_i + h_i)}$
\\ \hline  
$Q(t_0, {\bf x})$ & 
$  Q_0 + \sum\limits_{i=1}^{3} q_{i} \cos{(m_i x_i + d_i)}
$
\\ [4pt] \hline
$K(t_0, {\bf x})$ & $-H_0>0$
   \\ \hline 
&\\   [-10pt]
\quad$Z_{ij}^{TT}(t_0, {\bf x})$ \,& 
 $\begin{psmallmatrix}
b_2 +  c_2 \cos{x_2} & {\;} & \xi & {\;} &  \kappa_1+ c_1 \cos{x_2}  \\
&&&&\\
\xi & \; &  b_1+  a_1  \cos{x_1} & {\;} &  \kappa_2 +a_2 \cos{x_1}  \\
&&&&\\
\kappa_1+ c_1 \cos{x_2} 
&  {\;} &  \kappa_2+ a_2 \cos{x_1}  &  {\;} & -b_1 - b_2 - a_1 \cos{x_1} - c_2 \cos{x_2}
\end{psmallmatrix}$
\\ [14pt] \hline
 $Z_{ij}^L(t_0, {\bf x})$ &  fixed by the momentum constraint~\eqref{Cc-conf}   \\ \hline
   $\psi(t_0, {\bf x})$ & fixed by the Hamiltonian constraint~\eqref{Cg-conf}
   \\ \hline  
\end{tabular}
\caption{{\bf Initial conditions in robustness studies of slow contraction determined by fixing 30 or more parameters.} 
The scalar field $\phi$, its velocity $Q$, the mean curvature $K$, the transverse, traceless component $Z_{ij}^{TT}$ and the longitudinal component $Z_{ij}^L$ of the shear tensor $Z_{ij}$ as well as the conformal factor $\psi$ are specified at some initial time $t_0$ by fixing $\{\phi_0, f_i, n_i, h_i, Q_0, q_i, m_i, d_i, H_0, a_j, b_j, c_j, \kappa_j, \xi\}$, $(i=1,2,3; j=1,2)$.  
\label{table-1}}
\end{table*}

The York method is also employed in numerical relativity studies of inflation, albeit in a markedly  different way to date. A key difference is that initial conditions are restricted to those that trivially satisfy the momentum constraint~\eqref{Cc} by setting the scalar field's initial momentum density to be zero which is equivalent to imposing slow-roll conditions right at the onset rather than having slow-roll arise from more general initial conditions. 

This and other restrictions, such as a spatially uniform mean curvature or the absence of initial shear,  leave only 6 or less independent parameters for the initial conditions space of inflationary numerical relativity studies to date, as detailed in Table~\ref{table-2}. Note that all these restrictions are favorable to inflation. In addition, none of the studies to date has shown that non-linear effects can excite modes which enable exploration of generic conditions.
As illustrated in Ref.~\cite{Clough:2017efm}, in the only study to date in which a special form of small-amplitude shear perturbations is included,  the number of $e$-folds that could be reached is significantly limited.  
\begin{table*}[tbp]
\centering
\renewcommand{\arraystretch}{1.75}
\begin{tabular}{ |c| c|c|c|c|  }
 \hline
 \multirow{4}{*}{{\small Ref.}}  &  
{\footnotesize East\,et\,al\,\cite{East:2015ggf}} 
& \multirow{2}{*}{{\footnotesize Clough\,et\,al}} 
& \multirow{2}{*}{{\footnotesize Clough\,et\,al}} 
& \multirow{2}{*}{{\footnotesize Joana/Clesse}}
\\[-8pt]
& {\footnotesize Clough\,et\,al\,\cite{Clough:2016ymm, Clough:2017efm}}
& \multirow{2}{*}{{\footnotesize \cite{Clough:2016ymm, Clough:2017efm} }}
&  \multirow{2}{*}{{\footnotesize \cite{Clough:2017efm} }}
&  \multirow{2}{*}{{\footnotesize \cite{Joana:2020rxm}}}
\\[-8pt]
& {\footnotesize Aurrekoetxea\,et\,al\,\cite{Aurrekoetxea:2019fhr}}&  &  & 
\\[-8pt]
& {\footnotesize Joana/Clesse\,\cite{Joana:2020rxm}}&  &  & 
\\ \hline  \hline
$\phi$    &
\multicolumn{3}{c|}{
$\phi_0 + \sum\limits_{m} \phi_1 \left(\cos{m x_1} + \cos{m x_2} + \cos{m x_3} \right) $
}&
$\phi_0 $ 
\\ [8pt]\hline
$ \dot{\phi}$ & 0& $-\frac23 C$ &  0& {\small by Eq.\,\eqref{Cg}}
\\ \hline
$K$ & $-\sqrt{\langle\rho\rangle}$ & $ -C(\phi - \langle \phi\rangle)  $  & $-\sqrt{\langle\rho + \frac32\psi^8A_{ij}A^{ij}\rangle}$ 
&$K_0$
\\ \hline
&&&&\\   [-10pt]
\;$A_{ij}^{TT}$ \,& 0 & 0 &
 $  \begin{psmallmatrix}
0 & {\,} & f(x_3) & {\,} & f(x_2)  \\
&&&&\\
f(x_3) & \, &  0 & {\,} &  f(x_1)  \\
&&&&\\
f(x_2) &  {\,} &  f(x_1) &  {\,} & 0
\end{psmallmatrix}$
& 0
\\[15pt] \hline
$A_{ij}^L$ &  0& 0& 0& 0
\\ \hline
$\psi$ & \multicolumn{3}{c|}{ {\small fixed by Eq.}~\eqref{Cg}} 
& $\sqrt{e^{ e^{ -{\bf x}^2/\sigma^2 }}}$
\\ \hline
\end{tabular}
\caption{{\bf Initial conditions in robustness studies of inflation to date determined by fixing only 6 or less parameters.}  The scalar field $\phi$, its velocity $\dot{\phi}$, the mean curvature $K$, the transverse, traceless component $A_{ij}^{TT}$ and the longitudinal component $A_{ij}^L$ of the shear tensor $A_{ij}$ as well as the conformal factor $\psi$  are specified at some initial time $t_0$ by fixing $\{\phi_0, \phi_1, C, M, A,m_A \}$, where $m, m_A$ are integers with $1\leq \frac{L}{2\pi}\,m, \frac{L}{2\pi}\,m_A \leq M$, $f(x_i) = A\cos{(m_A x_i)}$,  $\rho = \frac12 (\partial_i\phi)^2 + V(\phi)$, $\langle \;\rangle$ denote averaging over the
spatial volume and $K_0= \sqrt{24(\Psi + V(\phi_0))}$ with $\Psi$ being fixed through the conformal factor $\psi$ only as defined in Eq.~(44) of Ref.\,\cite{Joana:2020rxm}. 
\label{table-2}}
\end{table*}
%

{\it Spatial and temporal resolution.} 
The sinusoidal form of spatially varying terms in both Tables~\ref{table-1}~and~\ref{table-2} reflect that, in both types of studies, periodic boundary conditions have been chosen. In studies of both slow contraction and inflation, the box, which corresponds to the initial patch, has an edge length of $2\pi |H_0^{-1}|$, where $|H_0^{-1}|$ is the initial Hubble radius. However, in both scenarios, if the physical situation is captured properly, after only a few time steps, the physical size of the initial patch is exponentially large compared to the Hubble volume because physical length scales evolve at exponentially different rates than the Hubble scale, as detailed above in Sec.~\ref{sec:standard}. The different dynamics of the patch versus the Hubble volume  poses a two-fold problem: 
\begin{itemize}
\item how to run the simulations sufficiently long such that smoothing is established for the 60 or more $e$-folds required to explain the observed homogeneity, isotropy and flatness of the universe?
\item how to implement a spatial resolution such that it is possible to track smoothing over both the super-Hubble patch and a single Hubble volume?
\end{itemize}

In the case of slow contraction, the first issue is addressed by using Hubble-normalized variables. The change in the Hubble radius is tracked by the time coordinate that measures the number of $e$-folds of contraction of the Hubble radius. Since the Hubble radius is not a dynamical quantity of the numerical scheme, the simulation involves a single scale only, namely, the physical length scale, corresponding to edge length of the box, which hardly changes during slow contraction. Hence, numerical relativity simulations of slow contraction easily run several hundreds of $e$-folds, establishing sufficient smoothing without encountering stiffness issues. One might worry, though, that Hubble-normalization makes it impossible to establish smoothing at the level of the rapidly shrinking Hubble volume. But, in studies of slow contraction, this second problem does {\it not} arise in the first place due ultralocality. As illustrated above in Sec.~\ref{sec:ultralocal}, within only a few $e$-folds of contraction, spacetime points whether inside or outside the same Hubble volume evolve independently of one another. That is, well before the Hubble radius shrinks to a size smaller than the simulation grid size, the dynamics eliminates any scale other than the simulation box size corresponding to the physical volume of the initial patch.

In the case of inflation, neither of the two issues has been addressed to date. %
To truly test for robustness, future numerical relativity studies of inflation need to significantly expand the range of initial conditions to include those that strongly disfavor inflation. 
In addition, two challenges must be met: (1) improving the inflationary codes so they can be run for at least  60 $e$-folds {\it when} starting from general initial conditions; and (2) finding a proper method to simultaneously establish smoothing on both the super-Hubble scales of the physical volume and the rapidly shrinking Hubble scale.

\section{Future directions}

Introducing numerical relativity  as a new complementary tool in Fundamental Cosmology opens up many avenues for future research. Since this approach has only began to be utilized, the potential for true discoveries is realistic, as illustrated by the examples \cite{Cook:2020oaj,Ijjas:2020dws,Ijjas:2021gkf,Ijjas:2021wml} establishing ultralocality as the key to a novel, non-linear smoothing mechanism in slow contraction models. 

First, as an immediate task, tests of robustness are needed for inflation models that accept more general initial conditions including those that disfavor inflation, run at least 60 or more $e$-folds and resolve scales ranging from the size of the physical patch down to the Hubble volume. 

Second, numerical relativity can provide a tool to study the dynamics of many different scenarios within slow contraction and inflationary cosmology and investigate possible new non-linear effects that are not captured by conventional perturbative and/or effective field theory analyses. Since the non-linear, non-perturbative regime of cosmological models is yet unexplored, surprises are rather likely.
A recent example of an unexpected result is the numerical relativity study of a simple gravitational theory which involves two kinetically-coupled scalar fields \cite{Ijjas:2021zyf}. Surprisingly, non-linear gravitational effects lead to a different background behavior than expected from  conventional perturbative analysis, demonstrating that the FRW solution might not be the true stable attractor of the full theory even though it appears to be the stable attractor perturbatively.

Third, it is important -- especially in the light of many upcoming observations -- to study the non-linear dynamics of other early- and late-universe scenarios that involve modifications of gravity beyond Einstein Relativity, such as Galilean genesis \cite{Nicolis:2008in,Creminelli:2006xe,Creminelli:2016zwa,Ageeva:2020buc}, matter bounce models \cite{Brandenberger:2012zb} or dynamical Chern-Simons gravity \cite{Alexander:2009tp}. Elaborating dynamical effects of the coupled, non-linear PDE system that cannot be captured by the truncated ODEs of cosmological perturbation theory or effective field theory methods is essential to establish the viability of these theories as well as to extract observational predictions. A proper numerical relativity scheme for theories that involve modifications of gravity requires, in particular, developing a yet missing framework that unites insights from perturbation theory and effective field theory with well-posedness considerations analogous to described in Sec.~\ref{subsec:WP}; see also \cite{Cayuso:2017iqc,Papallo:2017qvl,Allwright:2018rut,Bernard:2019fjb,Kovacs:2020ywu,Cayuso:2020lca}.

Finally, these investigations are closely tied to classical resolutions of the singularity problem of General Relativity. Recently, several non-singular bounce scenarios and models have been proposed based on conventional particle-physics inspired ideas \cite{Easson:2011zy,Alexander:2014uaa,Ijjas:2016vtq,Ijjas:2017pei,Graham:2017hfr,Kolevatov:2017voe,Brandenberger:2020eyf,Agrawal:2020xek}. To establish the viability of these proposals and to distinguish among them, {\it e.g.} by extracting their observational predictions, numerical relativity simulations will likely play a key role. Using cosmology as a  ground for stress-testing the non-linear, non-perturbative regime of these theories ideally prepares for elaborating connections with black hole physics.
\newline
\\
{\it Acknowledgments.} Thanks  to G. Gabadadze, D. Garfinkle, L. Lehner, V. Mukhanov, F. Pretorius, and P.J. Steinhardt for very helpful discussions. This research is supported by the Simons Foundation grant number 663083.





\bibliographystyle{apsrev}
\bibliography{gravity-review}

\end{document}